\documentclass[aps,prd,superscriptaddress,twoside,twocolumn,nofootinbib,%
10pt,showpacs,floatfix]{revtex4-1}

\usepackage{amsmath,amssymb}
\usepackage{graphicx,bm}
\usepackage{slashed}
\usepackage{dcolumn}

\allowdisplaybreaks

\begin{document}


\title{Spin-1 diquark contributing to the formation of tetraquarks in light mesons}


\author{Hungchong Kim}%
\email{hungchong@kau.ac.kr}
\affiliation{Research Institute of Basic Science, Korea Aerospace University, Goyang, 412-791, Korea}

\author{Myung-Ki Cheoun}%
\affiliation{Department of Physics, Soongsil University, Seoul 156-743, Korea}

\author{K. S. Kim}
\affiliation{School of Liberal Arts and Science, Korea Aerospace University, Goyang, 412-791, Korea}

\date{\today}


\begin{abstract}

We apply a mixing framework to the light meson systems and examine
tetraquark possibility in the scalar channel.
In the diquark-antidiquark model,
a scalar diquark  is a compact object when its color and flavor structures are in ($\bar{\bm{3}}_c$,
$\bar{\bm{3}}_f$). Assuming that all the quarks are in an $S$-wave,
the spin-0 tetraquark formed out of this scalar diquark has only one spin
configuration, $|J,J_{12},J_{34}\rangle=|000\rangle$,
where $J$ is the spin of the tetraquark, $J_{12}$ the diquark spin, $J_{34}$ the antidiquark spin.
In this construction of the scalar tetraquark, we notice that another compact diquark with spin-1 in
($\bm{6}_c$, $\bar{\bm{3}}_f$) can be used although it is less compact than the scalar diquark.
The spin-0 tetraquark constructed from this vector diquark leads to the spin configuration
$|J,J_{12},J_{34}\rangle=|011\rangle$.  The two configurations,
$|000\rangle$ and $|011\rangle$, are found to mix strongly through
the color-spin interaction. The physical states can be identified with certain mixtures of the two configurations
which diagonalize the hyperfine masses of the color-spin interaction.
Matching these states to two scalar resonances $a_0(980), a_0(1450)$
or to $K^*_0(800), K^*_0(1430)$ depending on isospin channel,
we find that their mass splittings are qualitatively consistent with the hyperfine mass splittings which can support
their tetraquark structure.
To test our mixing scheme further, we also construct the tetraquarks for $J=1,J=2$ with the spin configurations,
$|111\rangle$ and $|211\rangle$, and discuss possible candidates in the physical spectrum.

\end{abstract}


\maketitle

\section{Introduction}

Recently, there have been lots of progresses in the study of the multiquark states
which normally refer to hadrons containing four or higher number of quarks.
Among multiquarks, tetraquarks are quite interesting as there have been several studies
suggesting plausible evidences for their existence especially for hadrons containing heavy quarks.
The hidden-charmed resonance, $X(3872)$, measured in the $B$-meson decays~\cite{Belle03,Aubert:2004zr, Choi:2011fc, Aaij:2013zoa}
as well as
the other resonances with similar masses,
$X(3823)$~\cite{Bhardwaj:2013rmw}, $X(3900)$~\cite{Xiao:2013iha},
$X(3940)$~\cite{Abe:2007jna}, may be the tetraquarks
with the flavor structure
$cq\bar{c}\bar{q}~ (q=u,d)$~\cite{Maiani:2004vq,Maiani:2014aja,Kim:2016tys,Zhao:2014qva}.
Very recently, the LHCb collaboration~\cite{Aaij:2016iza,Aaij:2016nsc} reported
$X(4140)$, $X(4274)$, $X(4500)$, $X(4700)$
measured in $J/\psi \phi$ structures from the decays, $B^+\rightarrow J/\psi\phi K^+$.
Among various interpretations for them, tetraquarks are one of the most promising scenarios to explain their nature.

The tetraquark possibility was also investigated in the $D$ or $B$-meson excited states.
In Ref.~\cite{Kim:2014ywa}, we discussed that most of the $D$ or $B$-meson excited states
currently listed in Particle Data Group (PDG)~\cite{PDG16}, especially
their mass spectrum, can be understood if they are viewed as tetraquarks
with the diquark-antidiquark form, $cq{\bar q}{\bar q}, (q=u,d,s)$.
Using  the color-spin interaction, we reproduced the mass splittings of the
resonances in the excited states of $D$ and $B$ mesons quite successfully.
Also our model provides interesting phenomenology related to decays of spin-1 mesons,
which seems to fit nicely with
experimental observation.  Based on its phenomenological success,
we made some predictions for the $D$ and $B$ mesons to be found in future.

If the existence of the tetraquarks in heavy quark sector is confirmed, then it is likely that they
can exist also in the light meson system composed of $u,d,s$ quarks.  This is because the binding
among quarks in hadrons is governed by the color force which, in principle, does not discriminate
against the quark flavors.
Indeed, Jaffe proposed back in 1970s that, based on diquark-antidiquark picture,
$a_0 (980)$,  $f_0 (980)$, $\sigma(600)$, and $K^*_0 (800)$ may be tetraquarks forming a nonet in flavor
space~\cite{Jaffe77a,Jaffe77b,Jaffe04,Jaffe:1999ze}.
The main feature of this model starts from the fact that the spin-0
diquark belonging to a color and flavor antitriplet,
($\bar{\bm{3}}_c$, $\bar{\bm{3}}_f$), is the most compact object among all the possible diquarks.
The spin-0 tetraquarks
can be constructed by combining the spin-0 diquarks with the corresponding antidiquarks.
This type of the four-quark picture is further supported
by the other calculations~\cite{MPPR04a,EFG09}
even though it is still confronted with a two-quark picture involving a $P$-wave excitation~\cite{Torn95}.

What we want to emphasize in this work is that the above diquark with ($J=0$, $\bar{\bm{3}}_c$, $\bar{\bm{3}}_f$)
is not a unique choice even though it is an optimal starting point
in constructing tetraquarks in the diquark-antidiquark approach.
An alternative way is to construct scalar tetraquarks by
facilitating the spin-1 diquark with the color and flavor structures ($\bm{6}_c, \bar{\bm{3}}_f$).
This spin-1 diquark is a less compact object than the spin-0 diquark but it is still the second most
compact object among all the possible diquarks~\cite{Jaffe:1999ze}.
If we take this possibility into account, we then have two ways to construct tetraquarks
with spin-0.  These two tetraquarks are expected to mix each other which may lead to interesting
phenomena in the meson spectroscopy.  Therefore, we explore possible consequences of the mixing
between the two states in the spin-0 tetraquarks.

To make our investigation succinct, we focus on the isovector ($I=1$) and isodoublet ($I=1/2$) channels first of all.
If the two states ought to mix, the physical states must be generated by the diagonalization among them,
which should appear as
doublets in the actual spectrum. The ones with lower masses can be identified $a_0 (980)$ in isovector channel,
$K^*_0 (800)$~\footnote{$K^*_0 (800)$ is usually referred as $\kappa$. Here we follow
the nomenclature used in PDG.} in the isodoublet channel.
Then, the others with higher masses must be found in the meson spectrum.
Indeed, there are $a_0 (1450)$ in the isovector channel and $K^*_0 (1430)$ in isodoublet
channel, which can be identified as the candidates for this mixing scenario. As we will
discuss below, the mixing is important to generate the huge mass
splittings, about 500 MeV and 740 MeV,
from $a_0 (980)$ and $K^*_0 (800)$ respectively.

In fact, this type of the mixing was also discussed in Ref.~\cite{Jaffe04, Black:1998wt}.  There, this mixing
is used in a way to explain why the lowest lying states in $0^+$ channel have quite small masses below 1 GeV
without investigating the other states with higher masses.
Also Black {\it et al.}~\cite{Black:1999yz} discussed a different mixing scenario to explain $a_0 (1450)$ and $K^*_0 (1430)$.
Their mixing is between a $P$-wave $q\bar{q}$ and $qq\bar{q}\bar{q}$. This is different from our approach
where the mixing is introduced between the four-quark states
with different color and spin configurations.

In addition, there are other approaches that can be found in literature.
Ref.~\cite{Boglione:2002vv} proposed a model that $a_0(980)$ and $a_0(1450)$ can be dynamically generated
from a single $\bar{q}q$ state. A kind of hybrid model was also proposed where $a_0 (1450)$ and $K^*_0 (1430)$
are viewed as the tetraquarks mixed with a glueball state~\cite{Maiani:2006rq}.

Our approach based on spin-1 diquark should accompany two more spin states for the tetraquarks, namely $J=1,2$.
Finding corresponding resonances in PDG can provide further supports of our model.
Using the color-spin interactions, we also estimate the mass splittings
of these members from the spin-0 tetraquarks and look for the candidates in PDG which
can fit to our scheme.

This paper is organized as follows.
In Sec.~\ref{sec:wf}, we present tetraquark wave functions that could be relevant for
the light-meson systems. The wave functions for flavor, color and spin spaces
will be constructed using either the scalar or the vector diquark.
In Sec.~\ref{sec:mf}, we introduce the color-spin interaction as well as the color-electric
interaction and provide formulas for the hyperfine masses and the color-electric masses.
In Sec.~\ref{sec:results}, we present our
results and discuss their implication in the light-meson spectroscopy.
We summarize in Sec.~\ref{sec:summary}.


\begin{table}
\centering
\begin{tabular}{c|c|c|c|c}  \hline\hline
$J^{P}$ &  Meson  & $I$ & Mass (MeV) & $\Gamma$ (MeV) \\
 \hline
      & $a_0 (980)$ & $ 1 $ & 980 &  50-100   \\
$0^+$ & $a_0 (1450)$ & $ 1 $ & 1474 &  265  \\
\cline{2-5}
      & $K_0^* (800)$  & $1/2$ & 682 &  547    \\
      & $K_0^* (1430)$ & $1/2$ & 1425 & 270  \\
      & $K_0^* (1950)$ & $1/2$ & 1945 & 201  \\
\hline
      & $a_1 (1260)$ & $ 1 $ & 1230 & 250-600  \\
$1^+$ & $a_1 (1640)$ & $ 1 $ & 1647 & 254  \\
\cline{2-5}
      & $K_1 (1270)$ & $1/2$ & 1272 & 90  \\
      & $K_1 (1400)$ & $1/2$ & 1403 & 172  \\
      & $K_1 (1650)$ & $1/2$ & 1650 & 150  \\
\hline
      & $a_2 (1320)$ & $ 1 $ & 1318.3 & 105  \\
$2^+$ & $a_2 (1700)$ & $ 1 $ & 1732 & 194  \\
\cline{2-5}
      & $K_2^* (1430)$ & $1/2$ & 1425 & 98.5  \\
      & $K_2^* (1980)$ & $1/2$ & 1973 & 373  \\
\hline\hline
\end{tabular}
\caption{Here we collect all the isovector ($a_0,a_1,a_2$) and
isodoublet ($K^*_0, K_1, K_2$) resonances with the positive parity from PDG
and arrange them according to their spins $J=0,1,2$.
We omit the other resonances like $f_0,f_1,f_2$ {\it etc} as they are not our concern in the present work.}
\label{resonances}
\end{table}

\section{Tetraquark wave functions}
\label{sec:wf}

In this section, we construct the four-quark wave functions which might be relevant for the
light mesons composed by $u,d,s$ quarks.
Our construction is based on the diquark-antidiquark picture with an assumption
that all the quarks are in an
$S$-wave state. This assumption constrains that the corresponding tetraquark candidates
must be sought in the resonances with the positive parity to begin with.
As possible candidates for them, we collect isovector and isodoublet resonances with $J^P=0^+,1^+,2^+$
in Table~\ref{resonances} from PDG.
In this work, we do not discuss the isoscalar resonances for simplicity.

In constructing tetraquarks, the well-known approach, as advocated by Jaffe, is to
facilitate the compact diquark, which
is in $J=0$ with the color antitriplet $\bar{\bm{3}}_c$ and
the flavor antitriplet $\bar{\bm{3}}_f$.  It may be worth mentioning
that, due to Pauli principle, the diquark must be in the spin state $J=0$ when its color and flavor
structures are fixed to $\bar{\bm{3}}_c$ and $\bar{\bm{3}}_f$.
The fact that this
diquark is the most compact object among all the possible diquarks can be
demonstrated straightforwardly by calculating the hyperfine mass
of the color-spin interaction~\cite{Jaffe:1999ze}.
Likewise, the tight antidiquarks should be in $J=0$ with $\bm{3}_c$, $\bm{3}_f$.

Combining the diquarks with the antidiquarks leads to the tetraquarks forming a nonet in flavor,
$\bar{\bm{3}}_f\otimes {\bm{3}}_f={\bm{8}}_f\oplus {\bm{1}}_f$.
The flavor structure of the tetraquarks, by adopting the tensor notation for multiplets,
can be expressed as
\begin{eqnarray}
[{\bf 8}_f]^i_{j} &=& T_{j}\bar{T}^{i}-\frac{1}{3} \delta^{i}_{j}T_{m}\bar{T}^{m}\ ,\\
{\bf 1}_f &=& \frac{1}{\sqrt{3}}T_{m}\bar{T}^{m}\ .
\end{eqnarray}
Here the diquark ($T_i$) and the antidiquark ($\bar{T}^i$) are represented by the quark flavors as
\begin{eqnarray}
T_i &=&\frac{1}{\sqrt{2}}\epsilon_{ijk}q_j q_k\equiv [q_j q_k]\ ,\nonumber\\
\bar{T}^i &=& \frac{1}{\sqrt{2}}\epsilon^{ijk}\bar{q}^j \bar{q}^k \equiv [{\bar q}^j {\bar q}^k] \ .
\end{eqnarray}

To avoid further complications coming from the mixing between the flavor octet and
singlet among the isoscalar members,
our discussion in this work focuses on the isovector and isodoublet members which can couple to
$a_0$ and $K^*_0$.
To be more precise, the charged octet members, $a_0^+$ and $K^{*+}_0$, will be considered as
they are located at the boundary of the weight diagram where the multiplicity is just one.
The flavor wave functions that can couple to $a_0^+$ and $K^{*+}_0$ respectively are
\begin{eqnarray}
[{\bf 8}_f]^1_{2}=[su][\bar{d}\bar{s}]\ ; \quad [{\bf 8}_f]^1_{3}=[ud][\bar{d}\bar{s}]
\label{flavor wf}\ .
\end{eqnarray}
With this four-quark approach, $a_0$ has the hidden strange component, $s\bar{s}$, while $K^*_0$ contains one strange quark.
The experimental mass ordering, $M(a_0)\ge M(K^*_0)$,
can be understood more easily from this tetraquark picture than from the two-quark picture.

As for the color part of the wave function, the diquark is in $\bar{\bm{3}}_c$, the
antidiquark is in $\bm{3}_c$, and the four-quark state in total must be colorless.
It means that, for each flavor combination involved in Eq.~(\ref{flavor wf}), if we call
the first two quarks as $q_1 q_2$, and the third and fourth antiquarks as $\bar{q}^3 \bar{q}^4$,
the four-quark system has the following color structure with the color normalization,
\begin{eqnarray}
\frac{1}{\sqrt{12}}  \varepsilon_{abd}^{} \ \varepsilon^{aef}
\Big ( q_1^b q_2^d \Big )
\Big ( \bar{q}^3_e \bar{q}^4_f \Big )\ .
\label{color wave function1}
\end{eqnarray}
Here the roman indices, $a,b,d,e,f$, denote the colors.
Since the diquark spin $J_{12}$ and the antidiquark spin $J_{34}$ are zero, the
total spin $J$ must be zero. Then, the spin structure for the tetraquarks of this type is
restricted to
\begin{equation}
|J,J_{12},J_{34}\rangle=|000\rangle_{\bar{\bm{3}}_c,\bm{3}_c}\ .
\label{spin0}
\end{equation}
Here the subscripts denote the color structures for the diquark and antidiquark.

Alternatively, other types of diquark are also possible in constructing the tetraquarks.
Considering only symmetry properties associated with the spin, color,
flavor of the two-fermion system,
it is possible to have other diquarks which have the structures,
($J=1,\bm{6}_c, \bar{\bm{3}}_f$), ($J=1,\bar{\bm{3}}_c, \bm{6}_f$), ($J=0,\bm{6}_c, \bm{6}_f$).
One can demonstrate through the color-spin interaction that
the first one with
($J=1,\bm{6}_c, \bar{\bm{3}}_f$) is the most attractive configuration among
these three~\cite{Jaffe:1999ze}.
In fact, other diquarks with the structures, ($J=1,\bar{\bm{3}}_c, \bm{6}_f$), ($J=0,\bm{6}_c, \bm{6}_f$), are not compact because the color-spin
interaction for them are repulsive.
Using the first one, one can construct another tetraquarks by combining the diquark with $(J=1,\bm{6}_c, \bar{\bm{3}}_f)$
and the antidiquark with $(J=1,\bar{\bm{6}}_c, \bm{3}_f)$.

The resulting tetraquarks form a nonet again in flavor. The octet members that can couple to $a^+_0, K^{*+}_0$,
have the same flavor
wave function as Eq.~(\ref{flavor wf}). But now the diquark is in $\bm{6}_c$ and the
antidiquark is in $\bar{\bm{6}}_c$ so that they can be combined into a color singlet.
Again, for each flavor combination involved in Eq.~(\ref{flavor wf}),
calling the first two quarks as $q_1 q_2$ and the rest two antiquarks as $\bar{q}^3 \bar{q}^4$,
the four-quark system has the following color structure
\begin{eqnarray}
\frac{1}{\sqrt{96}} \Big( q_1^a q_2^b+q_1^b q_2^a \Big )
\Big (\bar{q}^3_a \bar{q}^4_b+\bar{q}^3_b \bar{q}^4_a\Big )\ .
\label{color wave function2}
\end{eqnarray}
Here again the roman indices, $a,b$, denote the colors.

However, with this spin-1 diquark scenario, there are three possible spin states for tetraquarks.
Namely, tetraquarks have the spins $J=0,1,2$ with the following configurations,
\begin{eqnarray}
|011\rangle_{\bm{6}_c,\bar{\bm{6}}_c}\ ;\quad
|111\rangle_{\bm{6}_c,\bar{\bm{6}}_c}\ ;\quad
|211\rangle_{\bm{6}_c,\bar{\bm{6}}_c}\label{spin1}\ .
\end{eqnarray}
What is interesting is that the tetraquarks in the scalar channel, $|011\rangle_{\bm{6}_c,\bar{\bm{6}}_c}$, can mix with
Eq.~(\ref{spin0}) through the color-spin interaction. The hyperfine masses, which are expectation
values of the color-spin interaction, form a $2\times 2$ matrix in the basis, $|000\rangle,|011\rangle$.
A diagonalization is necessary in order to identify
the physical states in this scalar channel.
Therefore, if this framework is realized in the real world, there should be two resonances in the scalar mesons
for each member in the octet, Eq.~(\ref{flavor wf}).
Indeed, as shown in Table~\ref{resonances}, there are two isovector resonances, $a_0(980)$ and $a_0(1450)$.
Also, in isodoublet channel, there are three resonances, $K^*_0(800), K^*_0(1430), K^*_0(1950)$, and
two of them might be the candidates fitting to our framework.
In this sense, the situation is quite promising and it is worth pursuing the consequences of this scenario further.

If our expectation works, additional resonances can be anticipated in the spin configurations,
$|111\rangle_{\bm{6}_c,\bar{\bm{6}}_c}$, $|211\rangle_{\bm{6}_c,\bar{\bm{6}}_c}$.
Alternatively, they can be hidden in the continuum of two-meson decays.
Anyway, as one can see in Table~\ref{resonances}, there are various resonances in $J=1,2$ and some of them
might be possible candidates of this scenario. Therefore,
it is also interesting to study which of them fits to this scheme.


\begin{table*}[t]
\centering
\begin{tabular}{c|l}  \hline\hline
$\langle J,J_{12},J_{34}| V | J,J_{12},J_{34}\rangle $
&  ~Corresponding formulas for one specific flavor combination, $q_1 q_2 {\bar q}^3 {\bar q}^4$ \\ \hline
$\langle 000|V_{CS}|000 \rangle $
& $\displaystyle 2 v_0^{} \left [ \frac{1}{m_1^{} m_2^{}} + \frac{1}{m_3^{} m_4^{}}
\right ] $ \\[3mm]
$\langle 011|V_{CS}|011 \rangle $
& $\displaystyle \frac{v_0}{3}  \left [ \frac{1}{m_1^{} m_2^{}}+\frac{1}{m_3^{} m_4^{}} + \frac{5}{m_1^{} m_3^{}}
+\frac{5}{m_1^{} m_4^{}} + \frac{5}{m_2^{} m_3^{}} + \frac{5}{m_2^{} m_4^{}}
\right ] $  \\[3mm]
mixing, $\langle 000|V_{CS}|011 \rangle $
& $ \displaystyle \sqrt{\frac{3}{2}} v_0^{} \left [ \frac{1}{m_1^{} m_3^{}} + \frac{1}{m_1^{} m_4^{}}
+ \frac{1}{m_2^{} m_3^{}} + \frac{1}{m_2^{} m_4^{}} \right ] $ \\[3mm] \hline
$\langle 111|V_{CS}|111 \rangle $
& $\displaystyle \frac{v_0}{6} \left [ \frac{2}{m_1^{} m_2^{}} + \frac{2}{m_3^{} m_4^{}} + \frac{5}{m_1^{} m_3^{}}
+\frac{5}{m_1^{} m_4^{}} + \frac{5}{m_2^{} m_3^{}} + \frac{5}{m_2^{} m_4^{}}
\right ] $  \\[3mm]
$\langle 211|V_{CS}|211 \rangle $
& $\displaystyle \frac{v_0}{6} \left [ \frac{2}{m_1^{} m_2^{}} + \frac{2}{m_3^{} m_4^{}} - \frac{5}{m_1^{} m_3^{}}
-\frac{5}{m_1^{} m_4^{}} - \frac{5}{m_2^{} m_3^{}} - \frac{5}{m_2^{} m_4^{}}
\right ] $  \\ [3mm]\hline
$\langle 000|V_{CE}|000 \rangle $
& $\displaystyle -\frac{8}{3} v_1^{} \left [ \frac{1}{m_1^{} m_2^{}} + \frac{1}{m_3^{} m_4^{}}
+ \frac{1}{2 m_1^{} m_3^{}} + \frac{1}{2 m_1^{} m_4^{}} + \frac{1}{2 m_2^{} m_3^{}} + \frac{1}{2 m_2^{} m_4^{}}
\right ] $ \\[3mm]
$\langle 011|V_{CE}|011 \rangle $
& $\displaystyle \frac{2}{3} v_1^{} \left [ \frac{2}{m_1^{} m_2^{}}+\frac{2}{m_3^{} m_4^{}} - \frac{5}{m_1^{} m_3^{}}
-\frac{5}{m_1^{} m_4^{}} - \frac{5}{m_2^{} m_3^{}} - \frac{5}{m_2^{} m_4^{}}
\right ] $  \\[3mm]
$\langle 111|V_{CE}|111 \rangle $
& $~ = \langle 011|V_{CE}|011 \rangle $  \\[1mm]
$\langle 211|V_{CE}|211 \rangle $
& $~ = \langle 011|V_{CE}|011 \rangle $  \\[1mm]
\hline\hline
\end{tabular}
\caption{Formulas for the expectation values of $V_{CS}$ and $V_{CE}$ are presented for a specific flavor combination, $q_1 q_2 {\bar q}^3 {\bar q}^4$,
with respect to the states indicated in the first column. Note that
the diquark and antidiquark are in the color states, $\bar{\bm{3}}_c,\bm{3}_c$, for the spin configuration, $|000\rangle$, and
they are in $\bm{6}_c,\bar{\bm{6}}_c$ for $|011\rangle, |111\rangle, |211\rangle$. }
\label{expectation values of V}
\end{table*}

\section{Mass formulas}
\label{sec:mf}

Normally a hadron mass can be calculated by adding constituent quark masses and the
expectation value of the potential, $V$, generated by summing over
all the pairs of quark-quark interaction.
In this sense,  the formula for a hadron mass ($M_H$) can be written schematically as
\begin{equation}
M_H=\sum_{i} m_i +\langle V \rangle\label{mass_f}\ ,
\end{equation}
where $m_i^{}$ the constituent mass of the $i$-th quark.
The quark-quark interaction can have two different sources, one-gluon exchange potential~\cite{DeRujula:1975qlm,Keren07,Silve92,GR81}
and the instanton-induced interaction~\cite{OT89,Oka:1990vx}.
A common feature of the two sources is the color-spin interaction ($V_{CS}$) which usually generates
the mass splittings among hadrons with different spins but with the same flavor content.
In particular, this interaction can explain the mass differences between
the octet and decuplet baryons as well as between the spin-1 and spin-0 meson
octets~\cite{Lipkin:1986dx,Lee:2009rt,Kim:2016tys,Kim:2014ywa}.
The instanton-induced interaction further provides the color-electric term ($V_{CE}$) and
the constant shift~\cite{OT89,Oka:1990vx}.
Taking the two sources into account, the potential can be effectively parameterized as
\begin{eqnarray}
V = v_0 \sum_{i < j}  \lambda_i \cdot \lambda_j \, \frac{J_i\cdot J_j}{m_i^{} m_j^{}}
+v_1^{} \sum_{i < j}  \frac{\lambda_i \cdot \lambda_j}{m_i^{} m_j^{}} + v_2\ .
\end{eqnarray}
Here $\lambda_i$ denotes the Gell-Mann matrix for the color SU(3), $J_i$ the spin.
The first and second terms are called color-spin and color-electric interactions
and we denote them as
\begin{eqnarray}
V_{CS} &=& v_0 \sum_{i < j}  \lambda_i \cdot \lambda_j \, \frac{J_i\cdot J_j}{m_i^{} m_j^{}}\label{CS}\ ,\\
V_{CE} &=& v_1 \sum_{i < j}  \frac{\lambda_i \cdot \lambda_j}{m_i^{} m_j^{}} \label{CE}\ .
\end{eqnarray}
The parameters $v_0, v_1$ represent
the strength of the color-spin and color-electric interactions which, in principle, need
to be fitted from the hadron masses.
The constant shift $v_2$ could be flavor-dependent in general.

The hadron masses of our concern can be formally calculated by Eq.~(\ref{mass_f}) using
the four states that we have introduced in Eq.~(\ref{spin0}) and Eq.~(\ref{spin1}).
As we have discussed in Ref.~\cite{Kim:2016tys}, fitting
all the parameters $v_0,v_1,v_2$ with only the hadron masses of concern here may be questionable
as to whether the same parameters can be used in other set of hadrons in general.
To reduce the ambiguity coming from the parameters,  we
focus on the mass splittings among hadrons of concern.

Then one can approximate that the mass splittings are generated by the interactions,
$V_{CS}$ and $V_{CE}$, through
\begin{equation}
\Delta M_H \approx \Delta \langle V_{CS} \rangle + \Delta \langle V_{CE} \rangle\ ,
\label{mass_splitting}
\end{equation}
if the differences are taken for the hadrons with the same flavor content.
Here the expectation values are taken with respect to the states introduced
in Eqs.~(\ref{spin0}), (\ref{spin1}), and their differences constitute the right-hand side.
It turns out that the right-hand side is dominated by the color-spin interaction, $V_{CS}$.
The color-electric interaction, $V_{CE}$, although it contributes differently to the masses of
$|000\rangle_{\bar{\bm{3}}_c,\bm{3}_c}$ and to the masses of the other category, $|011\rangle_{\bm{6}_c,\bar{\bm{6}}_c}$,
$|111\rangle_{\bm{6}_c,\bar{\bm{6}}_c}$, $|011\rangle_{\bm{6}_c,\bar{\bm{6}}_c}$,
its contribution to the mass splitting, $\Delta M_H$, is almost negligible as we will demonstrate below.
In addition, since $V_{CE}$ is independent of the spins, the mixing term
between the two states in the scalar channel,
$\langle 000 |V_{CE}|011\rangle$, is zero by the orthogonality of the spin
states~\footnote{For simplicity, we suppress the subscripts indicating the color structures from now on.}.
The constant shift $v_2$ cancels in the differences.

The expectation values, $\langle V_{CS} \rangle$ and $\langle V_{CE} \rangle$,
which we call hyperfine mass and color-electric mass respectively,
can be calculated straightforwardly.
We suggest the readers to refer Ref.~\cite{Kim:2014ywa} for
the technical details.  In Table~\ref{expectation values of V},
we present all the formulas for hyperfine and color-electric masses for the various spin configurations
with one specific flavor combination, $q_1 q_2 {\bar q}^3 {\bar q}^4$.
We also present the mixing term appearing in the scalar channel.

Note that the parameter $v_0$ has a negative value based on the analysis of the
baryon spectroscopy~\cite{Kim:2014ywa,Kim:2016tys}.
So from
the formulas provided in the scalar channel, one can see that the color-spin interaction, $V_{CS}$,
provides a fair amount of binding.  Of course, the actual binding from $V_{CS}$ must
take into account the mixing between the two states $|000\rangle$ and $|011\rangle$.
In the spin-1 and spin-2 channel, one can also see that $|111\rangle$ is more bound than $|211\rangle$
as far as the color-spin interaction is concern. This makes the $|211\rangle$ state heavier than the
$|111\rangle$ state which
is consistent with the general hierarchy observed in the mass spectrum in hadrons.
The contribution from the color-electric interaction is small due to
the small strength $v_1$ as we will see below.

The final expressions for the hyperfine mass, $\langle V_{CS}\rangle$, and the
color-electric mass, $\langle V_{CE}\rangle$, can
be obtained by including various flavor combinations involved in Eq.~(\ref{flavor wf}).
In particular, for the isovector channel
which can couple to $a^+_0$, $a^+_1$ or $a^+_2$ depending on its spin, the hyperfine mass
can be written schematically as
\begin{eqnarray}
\langle V_{CS} \rangle &=& \frac{1}{4}
\Big [ \langle V_{CS} \rangle_{su{\bar d}{\bar s}}
+\langle V_{CS} \rangle_{su{\bar s}{\bar d}}\nonumber \\
&&~+\langle V_{CS} \rangle_{us{\bar d}{\bar s}}
+\langle V_{CS} \rangle_{us{\bar s}{\bar d}} \Big ]\ ,
\label{isovector_hf}
\end{eqnarray}
where the specified flavor combination in the subscripts and the normalization in front follow from
Eq.~(\ref{flavor wf}).  Since the flavor structure are the same for all the spin states, $J=0,1,2$,
we have this type of flavor formula common for the three spin states.
The corresponding formula for the color-electric mass, $\langle V_{CE} \rangle$, can be obtained simply by
replacing the subscript $CS\rightarrow CE$.

For the isodoublet channel which can couple to $K^{*+}_0$, $K^+_1$ or $K^{*+}_2$ depending on its spin, we have the similar
formula but with different flavors as
\begin{eqnarray}
\langle V_{CS} \rangle &=& \frac{1}{4}
\Big [ \langle V_{CS} \rangle_{ud{\bar d}{\bar s}}
+\langle V_{CS} \rangle_{ud{\bar s}{\bar d}}\nonumber \\
&&~+\langle V_{CS} \rangle_{du{\bar d}{\bar s}}
+\langle V_{CS} \rangle_{du{\bar s}{\bar d}} \Big ]\ .
\label{isodoublet_hf}
\end{eqnarray}
Again, the corresponding formula for the color-electric mass, $\langle V_{CE} \rangle$, can be obtained by
replacing the subscript $CS\rightarrow CE$ in this equation.

\section{Results and discussion}
\label{sec:results}

Now we present and discuss the results for the mass splittings obtained from the
expectation value of the color-spin and color-electric formulas provided in
Table~\ref{expectation values of V}.  For our numerical calculations, first we need
to determine input parameters appearing in Table~\ref{expectation values of V}.
We take the standard values for the constituent quark masses $m_u = m_d = 330$~MeV, $m_s = 500$~MeV as
in our previous works~\cite{Kim:2014ywa,Kim:2016tys}.
For the strength $v_0$ of the color-spin interaction, we test
two possible choices. One choice is to use the value
determined from the $D$ meson excited states
studied within the tetraquark framework where $v_0$ is fixed from the mass spliting of $D_0^*(2318)-D_2^*(2463)$~\cite{Kim:2014ywa}.
This gives $v_0\sim (-192.9)^3$ MeV$^3$.
The other choice is to use the value determined from $\Delta -N$ mass difference,
which gives a slightly different value as $v_0\sim (-199.6)^3$ MeV$^3$~\cite{Kim:2014ywa,Kim:2016tys}.
Since our results turn out to depend strongly on this parameter, we present
the two results obtained by using the two values of this parameter. We call the first one as
``Theory I'' and the second one as ``Theory II''.
But for an illustration purpose, we discuss mainly with the results from ``Theory I''
but, in the final results, we will show the both calculations.

For the color-electric interaction, the strength $v_1$ can not be determined for example from
the mass splittings of the baryon octet and decuplet as the two multiplets have
the same color structure. For this, we take the value determined by $N,\Delta,\Lambda$ masses as inputs~\cite{Kim:2016tys}.
It gives $v_1 \sim (71.2)^3$ MeV$^3$.
This value should be regarded as a qualitative estimate as it can depend on how it is extracted.
Nevertheless, the contribution from the color-electric terms to our results are very small
so that our results are not sensitive to this particular choice.


\begin{table}[t]
\centering
\begin{tabular}{c|c|c}  \hline\hline
$\langle  V_{CS}  \rangle $, $\langle  V_{CE}  \rangle $
&  $I=1$ channel
&  $I=1/2$ channel \\ \hline
$\langle 000|V_{CS}|000 \rangle $ & -173.88 & -218.67  \\
$\langle 000|V_{CE}|000 \rangle $ & -23.8 & -29.29  \\ \hline
$\langle 011|V_{CS}|011 \rangle $ & -331.48 & -400.9 \\
$\langle 011|V_{CE}|011 \rangle $ & -24.57 & -29.29 \\ \hline
mixing, $\langle 000|V_{CS}|011 \rangle $ & -222.29 & -267.82 \\ \hline
$\langle 111|V_{CS}|111 \rangle $ & -180.23 & -218.67 \\
$\langle 111|V_{CE}|111 \rangle $ & -24.57 &-29.29   \\ \hline
$\langle 211|V_{CS}|211 \rangle $ & 122.27 & 145.78  \\
$\langle 211|V_{CE}|211 \rangle $ & -24.57 & -29.29  \\
\hline\hline
\end{tabular}
\caption{The numerical values for $\langle V_{CS}\rangle$ and $\langle V_{CE}\rangle$ are presented here
for the specified spin configurations. The $I=1$ channel can couple to $a_0$, $a_1$, $a_2$ and
the $I=1/2$ isodoublet channel can couple to $K^*_0$, $K_1$, $K^*_2$.  Here we present the results
with ``Theory I'' which uses the color-spin interaction parameter as $v_0=(-192.9)^3$ MeV$^3$.
All the numbers are given in MeV unit.}
\label{numerical values of V}
\end{table}

Having set all the parameters involved, we now discuss the numerical values for the hyperfine masses and
color-electric masses.
Table~\ref{numerical values of V} presents those masses calculated with respect to the specified spin
configurations using $v_0=(-192.9)^3$ MeV$^3$ (``Theory I'').
There are several interesting features to discuss about this result.

First, the hyperfine mass for $|011\rangle$ is more negative than the one for $|000\rangle$.
It is quite different from the usual expectation that the tetraquarks involving the spin-0 diquark
is more bound than the tetraquarks containing the spin-1 diquark.  This interesting
aspect can be understood if we examine carefully
the formulas for $\langle 000|V_{CS}|000 \rangle$ and $\langle 011|V_{CS}|011 \rangle$
given in Table~\ref{expectation values of V}.
The color-spin interaction in principle acts on all the pairs of quarks. For the $|000\rangle$ case, the calculated
hyperfine mass is proportional
to $\sim 1/m_1 m_2 + 1/m_3 m_4$ which means that the color-spin interaction
is nonzero only for two quarks in the diquark or for two antiquarks in the antidiquark.
There is no terms like $1/m_1 m_3$, $1/m_2 m_4$, indicating that the color-spin interaction acting on
any quark-antiquark pair is zero for the $|000\rangle$ state.
But for the $|011\rangle$ case, as one can see from the formula for $\langle 011|V_{CS}|011 \rangle$
in Table~\ref{expectation values of V},
there are nonzero contributions coming from quark-antiquark pairs in addition to those
from two quarks in the diquark and two antiquarks in the antidiquark.
This precisely makes the hyperfine mass of the $|011\rangle$ state more negative.

Secondly, we notice that the mixing term between the two states, $|000 \rangle$ and $|011 \rangle$, is quite large.
The mixing term in the isovector channel for example is about
$\langle 000|V_{CS}|011 \rangle \sim -222$ MeV.  Therefore, the two states, $|000 \rangle$ and $|011 \rangle$,
must mix strongly in making the physical states.

Additional thing that can be seen from Table~\ref{numerical values of V} is that the color-electric masses are quite
small. Moreover, their magnitudes are essentially the same for all the spin configurations.
Only exception is the element, $\langle 000|V_{CE}|000 \rangle$, in the isovector channel but its value
is different only slightly from other color-electric masses.
Therefore, the color-electric masses almost cancel in the mass differences and our results below based on
the mass splittings are almost independent of the color-electric interaction.
That is, as long as our analysis focuses on the mass splittings, we can safely approximate that
\begin{equation}
\Delta M_H \approx \Delta \langle V_{CS} \rangle\ .
\label{mass_splitting2}
\end{equation}

\subsection{Results on Isovector Channel}

Let us begin with a discussion on the isovector channel ($I=1$) which can couple to $a_0$,$a_1$,$a_2$.
Because of the mixing between the two states in spin-0,
we have a $2\times2$ matrix for the hyperfine masses $\langle V_{CS}\rangle$
with respect to the states
$|000 \rangle$ and $|011 \rangle$.  This matrix needs to be diagonalized in order to get
the physical hyperfine masses.
For the isovector channel with spin-0, the hyperfine mass matrix whose elements collected from
Table~\ref{numerical values of V}, and the matrix after the diagonalization are
\begin{eqnarray}
\begin{array}{c|lr}
& |000 \rangle & |011 \rangle \\
\hline
|000 \rangle & -173.9 & -222.3\\
|011 \rangle & -222.3 & -331.5
\end{array}
\quad
&\rightarrow&
\quad
\begin{array}{c|lr}
 & |0^{a_0}_A \rangle & |0^{a_0}_B \rangle \\
\hline
|0^{a_0}_A \rangle & -16.8 & 0.0\\
|0^{a_0}_B \rangle & 0.0  & -488.5
\nonumber\
\end{array}\ .
\end{eqnarray}
Here we denote the eigenstates as $|0^{a_0}_A \rangle$ and $|0^{a_0}_B \rangle$ with
the superscript $a_0$ indicating the resonance that they can couple to.
Note, the difference between the diagonal members, which is the key ingredient of our prediction,
is amplified from 157.6 MeV to 472 MeV.
This shows that the mass splitting between the physical states $|0^{a_0}_A\rangle$, $|0^{a_0}_B\rangle$
is strongly driven by the mixing in the spin-0 channel.
Note, the color-electric term $\langle V_{CE}\rangle$ only shifts the diagonal masses by almost the same amount.
Its contribution to the mass splittings therefore cancels and the gap, 472 MeV, is practically unchanged even with $\langle V_{CE}\rangle$.

The eigenstates $|0^{a_0}_A\rangle, |0^{a_0}_B\rangle$ are related to the original spin configurations through
\begin{eqnarray}
|0^{a_0}_A\rangle &=& -0.817 \mid 000 \rangle + 0.577 \mid 011 \rangle \nonumber\ ,\\
|0^{a_0}_B\rangle &=& 0.577 \mid 000 \rangle + 0.817 \mid 011 \rangle \label{a_mixing} \ .
\end{eqnarray}
This result is somewhat consistent with Black {\it et al.}~\cite{Black:1998wt} where this mixing is used in a different context.
Anyway, this indicates that the eigenstate $|0^{a_0}_A\rangle$ is in the state $|000 \rangle$ with the probability of 67 percent and
in the $|011 \rangle$ state of 33 percent.
It is interesting to see that the eigenstate with lower hyperfine mass, $|0^{a_0}_B\rangle$, are
in the $|011 \rangle$ state with higher probability of 67 percent.

It may be worth mentioning that our tetraquarks have a meson-meson component which is
either suppressed or enhanced depending on the states given in Eq.~(\ref{a_mixing}).
Our tetraquarks, schematically expressed by $q_1 q_2 {\bar q}^3 {\bar q}^4$, can have a component where
$q_1{\bar q}^3$ and $q_2 {\bar q}^4$ are separately combined into a color singlet as well as the other
component where those two pair are separately combined into a color octet. The first component
corresponds to the meson-meson component. One can work out this type of recombination from $|000\rangle$,
$|011\rangle$ and demonstrate that the meson-meson component is suppressed for $|0^{a_0}_A\rangle$ and enhanced
for $|0^{a_0}_B\rangle$.  We expect that this aspect can provide an interesting phenomenology
relating to the ``fall-apart'' decays of $|0^{a_0}_A\rangle$ and $|0^{a_0}_B\rangle$~\cite{kk_prep}.

According to Eq.~(\ref{mass_splitting2}), the mass difference between $|0^{a_0}_A\rangle$ and $|0^{a_0}_B\rangle$
can be written in terms of the hyperfine mass difference.
By calling the masses of $|0^{a_0}_A\rangle$ and $|0^{a_0}_B\rangle$ as $M_{0A}$ and $M_{0B}$ respectively,
the calculated mass difference, which constitutes the result from ``Theory I'',
is $M_{0A}-M_{0B}=-16.8-(-488.5)= 471.7$ MeV~\footnote{If we include the color-electric masses,
this value is changed to 471.9 MeV, which means that the contribution from $V_{CE}$ to the mass splitting is almost negligible.},
meaning that $|0^{a_0}_B\rangle$
has lower mass than $|0^{a_0}_A\rangle$ by 472 MeV.
This is indeed a huge separation in masses between the two states in spin-0.
This observation clearly leads us to identify the states $|0^{a_0}_A\rangle$ and $|0^{a_0}_B\rangle$
with the physical resonances
\begin{equation}
|0^{a_0}_A\rangle=a_0 (1450)\ ;\quad |0^{a_0}_B\rangle=a_0 (980)\ ,
\end{equation}
because the experimental mass difference of these two states,
$1474 - 980 =494$ MeV~\footnote{Note that the experimental mass of $a_0 (1450)$ is 1474 MeV
which is different from the number in the nomenclature of $a_0 (1450)$.}, is quite close to our result,
only 20 MeV higher.
Our calculation with the different parameter, $v_0=(-199.6)^3$ MeV$^3$,
namely the ``Theory II'' result,
gives $M_{0A}-M_{0B}=522.8$ MeV which is about 29 MeV higher than the experimental mass splitting.
Therefore, for $a_0 (1450)$ and $a_0 (980)$, our tetraquark formalism seems to work quite well.

To test our approach further, we look for a possible candidate which can fit to the $J=1$ resonance
with the configuration, $|111\rangle$.
As one can see in Table~\ref{resonances},
there are two candidates in PDG with spin-1, $a_1(1260)$ and $a_1(1640)$. Or another possibility is that
the $|111\rangle$ state might be hidden in the continuum of two-meson decays which is then too broad to be observed.
The hyperfine mass of $|111\rangle$ is $-180.23$ MeV as shown in Table~\ref{numerical values of V},
which is higher than the hyperfine mass of $|0^{a_0}_B\rangle$, $-488.5$ MeV,
but lower than that of $|0^{a_0}_A\rangle$, $-16.8$ MeV.  Applying this hierarchy to the mass spectrum, we
may identify the state $|111\rangle$ with $a_1 (1260)$. The other resonance $a_1(1640)$ certainly
does not fit into this hierarchy.

Denoting the mass of the state $|111\rangle$ as $M_1$,
its mass splittings from the spin-0 members, $|0^{a_0}_A\rangle$, $|0^{a_0}_B\rangle$, are
obtained from the hyperfine mass splittings as
\begin{eqnarray}
M_{1}-M_{0B}&=&-180.2-(-488.5)=308.3~{\rm MeV}\nonumber \ ,\\
M_{1}-M_{0A}&=&-180.2-(-16.8)=-163.4~{\rm MeV}\nonumber\ .
\end{eqnarray}
These numbers should be compared with the experimental mass splittings, $250$ MeV between $a_1(1260)$ and $a_0 (980)$,
and $-244$ MeV between $a_1(1260)$ and $a_0 (1450)$. The hyperfine mass splittings are off by $50\sim 80$ MeV
from the experimental splittings.  Although the agreement is not precise, the errors are within an acceptable range
if one takes into account the broad decay width of $a_1(1260)$, $\Gamma=250 - 600$ MeV.
Of course, this identification needs to be further examined in future from other properties such as its decay modes and so on.

For the spin-2 case, there are two candidates in Table~\ref{resonances}, $a_2(1320)$ and $a_2(1700)$,
and one of them can be identified with $|211\rangle$.
The hyperfine mass of $|211\rangle$
in Table~\ref{numerical values of V} is $122.27$ MeV which is higher
than any of the hyperfine masses for the states $|0^{a_0}_A\rangle$, $|0^{a_0}_B\rangle$, $|111\rangle$.
Thus, the corresponding resonance to $|211\rangle$ must be
heavier than those in spin-0 and spin-1. The resonance, $a_2(1700)$, fits into this criteria and
it can be identified with $|211\rangle$.
Denoting the mass for $|211\rangle$ with $M_2$, its mass splittings from the spin-0 and spin-1 states
estimated from the hyperfine mass splittings are
$M_2-M_{0B} = 611$ MeV, $M_2-M_{0A} = 138$ MeV, $M_2-M_1 = 303$ MeV.
The corresponding mass splittings based on their experimental masses
in Table~\ref{resonances} are
$752$ MeV, $258$ MeV, $502$ MeV respectively.  The mismatch is less than two hundred MeV or so.
Again, although the agreement is not precise, the trend in mass differences seems to match more or less.
Also taking into account the broad widths associated with the resonances involved, we
can claim that the mismatch is not enough to rule out our four-quark scheme.

Our results for $a_0, a_1, a_2$ are summarized in Table~\ref{splitting_a}.
There, we present our results for ``Theory I'' and ``Theory II'' in comparison with the experimental
mass splittings based on the identifications
$|0^{a_0}_B\rangle=a_0(980)$, $|0^{a_0}_A\rangle=a_0(1450)$, $|111\rangle=a_1(1260)$, $|211\rangle=a_2(1700)$.
Both results qualitatively agree with the experimental splittings.
Based on these results, we may conclude that the spin-1
diquark seems to play an important role in the formation of the tetraquarks in light mesons.


\begin{table}
\centering
\begin{tabular}{l|c|c|c}  \hline\hline
\multicolumn{4}{c}{$a_0,a_1,a_2$ channel} \\
\hline
Participating  & Expt.& \multicolumn{2}{c}{$\Delta \langle V_{CS} \rangle$ (MeV)} \\
\cline{3-4}
spin states               &  $\Delta M_{H}$ (MeV) &   Theory I &  Theory II \\
 \hline
$|0^{a_0}_B\rangle - |0^{a_0}_A\rangle$ &  494  & 471.7 & 522.8 \\[0.3mm]
\hline
$|111\rangle - |0^{a_0}_B\rangle$ & 250 & 308.3 & 341.7 \\[0.3mm]
$|111\rangle - |0^{a_0}_A\rangle$ & -244 & -163.4 & -181.1     \\[0.3mm]
\hline
$|211\rangle - |0^{a_0}_B\rangle$ & 752  & 610.8 & 677.0   \\[0.3mm]
$|211\rangle - |0^{a_0}_A\rangle$ & 258  & 138.1 & 154.2  \\[0.3mm]
$|211\rangle - |111\rangle$       & 502 & 302.5 & 335.3   \\
\hline\hline
\end{tabular}
\caption{The hyperfine mass splittings among the spin states are compared with the corresponding resonances
in $a_0,a_1,a_2$ channel.  Here, we identify
$|0^{a_0}_B\rangle=a_0(980)$, $|0^{a_0}_A\rangle=a_0(1450)$, $|111\rangle=a_1(1260)$, $|211\rangle=a_2(1700)$.
The column under the name ``Theory I'' [``Theory II''] is obtained with the strength
 $v_0=(-192.9)^3$ MeV$^3$ [$v_0=(-199.6)^3$ MeV$^3$].  See the text for the choice of this parameter.
}
\label{splitting_a}
\end{table}

\subsection{Results on Isodoublet Channel}

We now move to a discussion for the isodoublet ($I=1/2$) channels which
can couple to $K^*_0$, $K_1$, $K^*_2$ resonances.
Again in the spin-0 case, because of the mixing,
we have a $2\times2$ matrix for the hyperfine masses $\langle V_{CS}\rangle$
with respect to the spin configurations $|000 \rangle$ and $|011 \rangle$.
The diagonalization leads to
\begin{eqnarray}
\begin{array}{c|lr}
& |000 \rangle & |011 \rangle \\
\hline
|000 \rangle & -218.7 & -267.8\\
|011 \rangle & -267.8 & -400.9
\end{array}
\quad
&\rightarrow&
\quad
\begin{array}{c|lr}
 & |0^{K_0}_A \rangle & |0^{K_0}_B \rangle \\
\hline
|0^{K_0}_A \rangle & -26.9 & 0.0\\[0.3mm]
|0^{K_0}_B \rangle & 0.0  & -592.7
\nonumber
\end{array}\ .
\end{eqnarray}
Here we have introduced the superscript $K_0$ in the eigenstates to indicate that the states can couple to $K^*_0$.
The eigenstates $|0^{K_0}_A\rangle, |0^{K_0}_B\rangle$ are related to the original spin configurations through
\begin{eqnarray}
|0^{K_0}_A\rangle &=& -0.813 \mid 000 \rangle + 0.582 \mid 011 \rangle \nonumber\ ,\\
|0^{K_0}_B\rangle &=& 0.582 \mid 000 \rangle + 0.813 \mid 011 \rangle \label{K_mixing} \ .
\end{eqnarray}
The mixing parameters are not so different
from the isovector case, Eq.~(\ref{a_mixing}).

We observe again that the mixing drives a huge separation of the diagonal hyperfine masses, about 565.8 MeV.
The eigenstates $|0^{K_0}_A\rangle$ and $|0^{K_0}_B\rangle$ need to be identified with the physical resonances.
Among three possible candidates with spin-0 in Table~\ref{resonances},
$K^*_0(800)$, $K^*_0(1430)$, $K^*_0(1950)$, it may be appropriate to take
the two states with lower masses, i.e.,
\begin{equation}
|0^{K_0}_A\rangle=K^*_0 (1430)\ ;\quad |0^{K_0}_B\rangle=K^*_0 (800)\ .
\end{equation}
Using the experimental masses in PDG for $K^*_0 (1430)$ and  $K^*_0 (800)$, their mass difference is
$\Delta M_{H}=1425-682=743$ MeV, which is
higher than the hyperfine mass splitting of 565.8 MeV.
Considering the fact that the decay widths of $K^*_0 (800)$ and $K^*_0 (1430)$  are $547$ MeV, $270$ MeV respectively,
we may claim that our mixing scheme qualitatively works for this spin-0 isodoublet channel.

In the spin-1 case, there are three possible candidates in Table~\ref{resonances}, $K_1 (1270)$, $K_1 (1400)$, $K_1 (1650)$
and one of them can be matched with our spin state $|111\rangle$.
We choose one resonance by looking at the mass hierarchy generated from the hyperfine masses.
The hyperfine mass for the state $|111\rangle$ is $-218.7$ MeV as can be seen in Table~\ref{numerical values of V}.
Comparing this with the hyperfine masses for $|0^{K_0}_A\rangle$, $|0^{K_0}_B\rangle$, one can establish
the mass hierarchy as
$|0^{K_0}_A\rangle > |111\rangle > |0^{K_0}_B\rangle$.
The resonance $K_1 (1270)$ fits to this hierarchy relatively well. The other candidate $K_1(1400)$,
although it barely fits to the hierarchy, its mass gap from $K^*_0 (1430)$ seems too narrow.
With this identification, its mass splittings from the spin-0 resonances agree at least qualitatively
with the hyperfine mass splittings as one can see in the second and third line from the top in Table~\ref{splitting_K}.

A somewhat puzzling situation occurs for the spin-2 case. In Table~\ref{resonances},
there are two candidates, $K^*_2(1430)$, $K^*_2(1980)$, that can be
matched with the spin state, $|211\rangle$. According to Table~\ref{numerical values of V},
the hyperfine mass for $|211\rangle$ is $145.8$ MeV, which is $502$ MeV higher than the
hyperfine mass of $|111\rangle$. With the identification with $|111\rangle=K_1 (1270)$,
we need to have a spin-2 resonance with a mass around 1770 MeV.  But
the mass of $K^*_2(1430)$ is too small and the mass of $K^*_2(1980)$ is too large.
It is somewhat hesitating to identify either of the resonances as
$|211\rangle$ even if we take into
account the broad width associated with the resonances.
Nevertheless, by identifying $|211\rangle=K^*_2(1430)$, we obtain
the experimental mass splittings,
$M_2-M_{0B}= 743$ MeV,  $M_2-M_{0A}= 0$ MeV, $M_2-M_{1}= 153$ MeV.
The first number is consistent with
our calculation but the second and third ones seems a little too far to fit our results
given under ``Theory I'' and ``Theory II'' in Table~\ref{splitting_K}.
If we identify $|211\rangle=K^*_2(1980)$ instead, the experimental mass splittings
associated with the spin-2 resonance
become $M_2-M_{0B}= 1291$ MeV, $M_2-M_{0A}= 548$ MeV, $M_2-M_{1}= 701$ MeV, which do not
fit to our calculation also.

There could be various reasons for the disagreement in the spin-2 case.
It is possible that the corresponding candidate may be hidden
in two-meson continuum
or has not be observed yet. Alternatively there might be
some other mechanisms, such as configuration mixing with different multiplets, to change the mass of the
spin-2 resonance in the isodoublet.
Anyway, it would be interesting to investigate this problem further in future.


\begin{table}
\centering
\begin{tabular}{l|c|c|c}  \hline\hline
\multicolumn{4}{c}{$K^*_0,K_1,K^*_2$ channel } \\
\hline
Participating  & Expt. & \multicolumn{2}{c}{$\Delta \langle V_{CS} \rangle$ (MeV)} \\
\cline{3-4}
spin states    & $\Delta M_{H}$ (MeV)    &   Theory I &  Theory II  \\
 \hline
$|0^{K_0}_B\rangle - |0^{K_0}_A\rangle$ &  {743}  & 565.8 & 627.1 \\[0.3mm]
\hline
$|111\rangle - |0^{K_0}_B\rangle$ &  {590} & 374.0 & 414.5   \\[0.3mm]
$|111\rangle - |0^{K_0}_A\rangle$ &  -153 & -191.8 & -212.6  \\[0.3mm]
\hline
$|211\rangle - |0^{K_0}_B\rangle$ &  743 & 738.5 & 818.5   \\[0.3mm]
$|211\rangle - |0^{K_0}_A\rangle$ &  0 & 172.7 & 191.4  \\[0.3mm]
$|211\rangle - |111\rangle$ &  153 & 364.5 & 403.9   \\
\hline\hline
\end{tabular}
\caption{The hyperfine mass splittings among the spin states are compared with the corresponding resonances
in $K^*_0,K_1,K^*_2$ channel.
Here, we identify $|0^{K_0}_B\rangle=K^*_0(800)$, $|0^{K_0}_A\rangle=K^*_0(1439)$, $|111\rangle=K_1(1270)$, $|211\rangle=K^*_2(1430)$.
The spin-2 resonance seems not fit into our tetraquark framework.
For the other explanations for this table, see the caption of Table~\ref{splitting_a}.}
\label{splitting_K}
\end{table}

\section{Summary}
\label{sec:summary}

In this work, we have proposed two possible ways to construct tetraquarks in the light-meson system.
The standard way is to facilitate the spin-0 diquark and spin-0 antidiquark to form a flavor nonet.
In this approach, the color and flavor structures for the diquark are ($\bar{\bm{3}}_c$,$\bar{\bm{3}}_f$)
and for the antidiquark, they are ($\bm{3}_c$,$\bm{3}_f$).
The tetraquarks formed in this way has one spin configuration only, $|J,J_{12},J_{34}\rangle=|000\rangle$.
The other way to construct tetraquarks is to facilitate the spin-1 diquark and antidiquark where the diquark
is in ($\bm{6}_c$,$\bar{\bm{3}}_f$) and the antidiquark is in ($\bar{\bm{6}}_c$, $\bm{3}_f$).
This construction is motivated by the fact that the spin-1 diquark with ($\bm{6}_c$,$\bar{\bm{3}}_f$)
is the second most attractive among all the possible diquarks.
With this approach, the tetraquarks can have three spin states with the
configurations, $|011\rangle$, $|111\rangle$, $|211\rangle$.

Therefore, for spin-0 tetraquarks, there are two spin configurations, $|000\rangle$ and $|011\rangle$,
and they are found to mix strongly through the color-spin interaction.
We have found that the physical states obtained from the diagonalization of the hyperfine mass matrix
match qualitatively well to $[a_0(980), a_0(1450)]$ in the hidden strangeness channel and
$[K^*_0(800), K^*_0(1430)]$ in the open strangeness channel.

To solidify our tetraquark framework, we have also looked for physical resonances that can be matched
to the additional states $|111\rangle$ and $|211\rangle$.
Our analysis from the mass splittings suggests that $a_1(1260)$ and $K_1(1270)$ may be the candidates
for $|111\rangle$ and $a_2(1700)$ could be a candidate for $|211\rangle$.
But there is one resonance seemingly missing in spin-2 with open strangeness channel as
neither of the existing resonances $K_2 (1430)$ and $K^*_2(1980)$ in that channel
seems to fit into our framework.
Nevertheless, based on qualitative agreement in most spin channels, we believe that
our tetraquark formalism may be realized in the light-meson system.
Further studies such as their decay pattern and so on are necessary in order to establish this model.

\acknowledgments

\newblock
The work of H.Kim was supported by Basic Science Research Program through the National Research Foundation of Korea(NRF)
funded by the Ministry of Education(Grant No. 2015R1D1A1A01059529).
The work of K.S.Kim was supported by the National Research Foundation of Korea (Grant No. 2015R1A2A2A01004727).

\end{document}